\documentclass[runningheads]{llncs}
\usepackage{hyperref}
\usepackage{subcaption}
\usepackage{tabularx}
\usepackage{listings}
\usepackage{verbatim}
\usepackage[table]{xcolor}
\usepackage{setspace}
\usepackage{enumitem}
\usepackage{float}
\usepackage[]{caption}
\usepackage{orcidlink}
\usepackage{pdfpages}
\usepackage{enumitem}
\usepackage{graphicx}
\usepackage{booktabs}
\usepackage{amssymb}
\usepackage{bbm}
\usepackage{array}
\begin{document}

\title{Challenges of Anomaly Detection in the Object-Centric Setting: Dimensions and the Role of Domain Knowledge}
\titlerunning{Challenges of Anomaly Detection in the Object-Centric Setting}

\author{
Alessandro Berti\inst{1,2}\orcidlink{0000-0002-3279-4795},
Urszula Jessen\inst{3,4}\orcidlink{0000-0002-7282-8451},
Wil M.P. van der Aalst\inst{1,2}\orcidlink{0000-0002-0955-6940},
Dirk Fahland\inst{4}\orcidlink{0000-0002-1993-9363}
}
\authorrunning{A. Berti, U. Jessen, et al.}
\institute{Process and Data Science Chair, RWTH Aachen University, Aachen, Germany \and
Fraunhofer FIT, Sankt Augustin, Germany \and
Process Insights, ECE Group Services, Hamburg, Germany \and
Eindhoven University of Technology, The Netherlands \\
\email{\{a.berti, wvdaalst\}@pads.rwth-aachen.de; \{u.a.jessen, d.fahland\}@tue.nl}}
\maketitle

\begin{abstract}
Object-centric event logs, allowing events related to different objects of different object types, represent naturally the execution of business processes, such as ERP (O2C and P2P) and CRM. However, modeling such complex information requires novel process mining techniques and might result in complex sets of constraints. Object-centric anomaly detection exploits both the lifecycle and the interactions between the different objects. Therefore, anomalous patterns are proposed to the user without requiring the definition of object-centric process models. This paper proposes different methodologies for object-centric anomaly detection and discusses the role of domain knowledge for these methodologies. We discuss the advantages and limitations of Large Language Models (LLMs) in the provision of such domain knowledge. Following our experience in a real-life P2P process, we also discuss the role of algorithms (dimensionality reduction+anomaly detection), suggest some pre-processing steps, and discuss the role of feature propagation.
\keywords{Object-Centric Anomaly Detection \and Object-Centric Feature Extraction \and Procurement Processes \and Large Language Models}
\end{abstract}

\renewcommand{\sectionautorefname}{Section}
\renewcommand{\subsectionautorefname}{Section}
\renewcommand{\subsubsectionautorefname}{Section}
\def\univs{U}
\newcommand{\univ}[1]{\univs_{\mathit{#1}}}
\newcommand{\class}[1]{\mathbb{C}_{\mathit{#1}}}
\newcommand{\pim}[1]{\pi_{\mathit{#1}}}

% Redefine the definition environment to include a smaller font size
\let\olddefinition\definition
\renewcommand{\definition}{\small\olddefinition}

\lstset{
  basicstyle=\tiny,
  columns=fullflexible,
  frame=single,
  breaklines=true,
  %postbreak=\mbox{\textcolor{red}{$\hookrightarrow$}\space},
}

\section{Introduction}
\label{sec:introduction}

Process mining, a branch of data science, derives insights from data recorded by information systems on business processes. Traditional techniques assume each event is linked to a single case, causing repeated data extractions, minimal consideration of the interactions between different object types, and issues like deficiency (events excluded if they don't fit the case notion), convergence (events related to several objects of the same object type are replicated in different cases), and divergence (events in a case having the same activity may be related to different objects, leading to misleading causalities) \cite{DBLP:conf/sefm/Aalst19}. Object-Centric Process Mining (OCPM) eliminates the single-case assumption, accommodating events involving various object types.
Several techniques for process discovery \cite{DBLP:journals/fuin/AalstB20} and conformance checking \cite{DBLP:conf/bis/FahlandLDA11,DBLP:conf/icpm/ParkA22} have been proposed in the object-centric setting. In practice, the main challenge is the large number of object types in real-life logs. For example, a Purchase-to-Pay (P2P) log includes multiple object types such as purchase requisitions, orders, invoices, and payments, and their interactions. It is challenging to define process models or constraints that cover the lifecycle and interactions of numerous object types. Additionally, assessing results from mainstream conformance checking techniques is difficult \cite{DBLP:conf/s-bpm-one/DunzerSMB19}, with techniques often focusing on just one perspective: control-flow, time or data.

\emph{Anomaly detection} \cite{DBLP:journals/jbd/ThudumuBJS20} is the process of identifying data points, events, or observations that deviate significantly from the expected pattern within a dataset. This technique is essential in various fields, including fraud detection, network security, fault detection, and medical diagnostics.
\emph{Anomaly detection in the OCPM setting} focuses on identifying business objects that behave differently from others. By analyzing interactions, anomalies such as discrepancies in quantities and amounts between invoices and orders can be detected. Additionally, integrating temporal data enables anomaly detection based on workload variations. Challenges arise when anomaly detection spans multiple features and interactions among different object types. The computational complexity increases with the addition of features, and the complexity of interpreting results manually becomes substantial.

\begin{figure*}[!t]
\vspace{-3mm}
\centering
\includegraphics[width=0.8\textwidth]{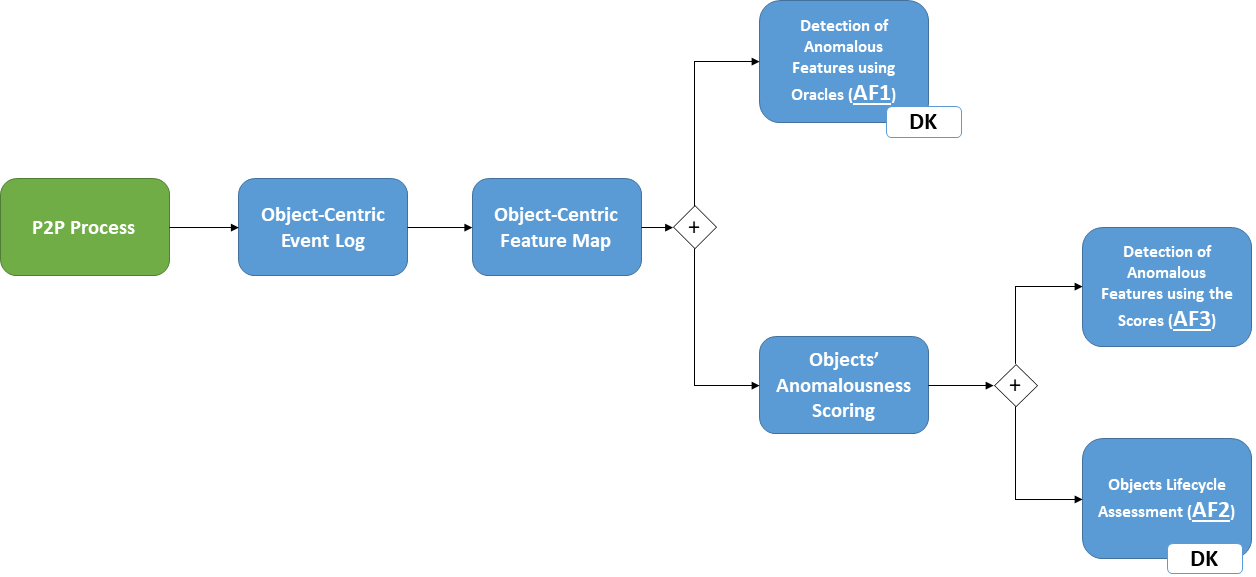}
\caption{Outline of the contributions proposed in the paper. The approaches highlighted with ``DK'' require domain knowledge.}
\label{fig:introductionOutline}
\vspace{-6mm}
\end{figure*}

In this paper, we discuss challenges in object-centric anomaly detection, focusing on methodologies for obtaining explainable and actionable insights. We introduce techniques that use domain knowledge to identify \emph{contextual outliers} and methods for interpreting anomalies without domain knowledge (\emph{structural outliers}). This includes a review of prevalent anomaly detection and feature selection methods. Our analysis used a real-life P2P process event log, employing the \emph{pm4py} process mining library and the \emph{OC-PM} tool.
Figure \ref{fig:introductionOutline} summarizes the paper's contributions. We explore methods to identify anomalous features:
\vspace{-2mm}
\begin{enumerate}[label=AF\arabic*]
\item We introduce an \emph{oracle} that evaluates features with their numerical values to determine anomalies. This oracle utilizes domain knowledge\footnote{It can be either a human analyst or an LLM}.
\item An anomaly detection algorithm assigns scores to objects, allowing a domain knowledge owner to examine the lifecycles of objects with the lowest scores for detailed anomaly patterns analysis.
\item The anomaly scores of objects can be aggregated to generate a feature-level anomaly score.
\end{enumerate}
\vspace{-2mm}

The paper is structured as follows:
Section \ref{sec:relatedWork} reviews related work.
Section \ref{sec:preliminaries} defines key concepts for object-centric feature extraction.
Section \ref{sec:approach} outlines methods for detecting anomalous object-centric features.
Section \ref{sec:discussion} discusses a case study.
Section \ref{sec:conclusion} concludes the paper.

\vspace{-2mm}
\section{Related Work}
\label{sec:relatedWork}

\emph{Feature Extraction and Anomaly Detection}: Fundamental work on feature extraction and machine learning for traditional event logs is discussed in \cite{DBLP:journals/is/LeoniAD16}. In the object-centric context, graph-based feature extraction from event logs is detailed in \cite{berti2022graph}, identifying various P2P process issues such as maintenance contracts and maverick buying. These features, integrated into \emph{pm4py} \cite{DBLP:journals/simpa/BertiZS23} and \emph{OC-PM} \cite{DBLP:journals/sttt/BertiA23}, support various machine learning algorithms. 
 The approach described in \cite{berti2023analyzing} uses these features for anomaly detection among other applications. Anomaly detection methods in process mining are reviewed in \cite{DBLP:journals/bise/KoC23}, with \cite{DBLP:conf/bpm/CasparyRA23} utilizing Large Language Models for semantic anomaly detection, and \cite{DBLP:journals/is/BohmerR20} focusing on identifying and explaining anomalies' root causes. \\
 \emph{Object-Centric Conformance Checking}: The approach described in \cite{DBLP:conf/bis/FahlandLDA11} divides object-centric conformance checking into lifecycle-based and interaction-based analysis. In \cite{DBLP:conf/bis/LiCA17}, object-centric behavioral constraint models are introduced as declarative rules for activity behaviors and relationships, yet lacking a conformance checking approach. In \cite{DBLP:conf/icpm/ParkA22}, rules are verified on object lifecycles and interactions, but these rules must be manually defined without a discovery method. Object-centric Petri nets for conformance checking are discussed in \cite{DBLP:journals/fuin/AalstB20}. In particular, \cite{DBLP:conf/er/LissAA23} uses object-centric alignments to match event log behavior with a Petri net model, and \cite{DBLP:conf/icpm/AdamsA21} defines fitness and precision in this context.

\section{Preliminaries}
\label{sec:preliminaries}

In this section, we present some of the basic concepts used in the rest of the paper.

\subsection{Object-Centric Event Logs}
\label{subsec:objectCentricEventLogs}

Object-centric event logs relax the assumption that an event is related to a single case notion.
Instead, an event can be related to different objects of different object types.

\begin{definition}[Universes]\label{def:universes}
We define the following universes:
$\univ{\Sigma}$ is the universe of strings (with ${{<}}_{\Sigma}$ being the lexicographic order);
$\univ{OT} \subseteq \univ{\Sigma}$ is the universe of object types;
$\univ{O} \subseteq \univ{\Sigma}$ is the universe of objects (identifiers);
$\univ{E} \subseteq \univ{\Sigma}$ is the universe of events (identifiers);
$\univ{act} \subseteq \univ{\Sigma}$ is the universe of activities (i.e., event types);
$\univ{att} \subseteq \univ{\Sigma}$ is the universe of attribute names;
$\univ{timest} \subseteq \mathbb{R}^{+}$ is the universe of timestamps;
$\univ{val}$ is the universe of attribute values.
\end{definition}

Definition \ref{def:ocel} presents the definition of object-centric event log, requiring the universes introduced in Definition \ref{def:universes}.

\begin{definition}[Object-Centric Event Log]\label{def:ocel}
An object-centric event log is a tuple $L = (E, \allowbreak O, \allowbreak \pi_{otyp}, \allowbreak \pi_{act}, \allowbreak \pi_{time}, \allowbreak \pi_{omap}, \allowbreak \pi_{vmap}, \allowbreak \pi_{ovmap}, \allowbreak {<})$ in which:
$E \subseteq \univ{E}$ is the set of events; $O \subseteq \univ{O}$ is the set of objects; $\pi_{otyp} : O \rightarrow \univ{OT}$ maps each object to an object type;
$\pi_{act} : E \rightarrow \univ{act}$ maps each event to an activity; $\pi_{time} : E \rightarrow \univ{timest}$ maps each event to a timestamp; $\pi_{omap} : E \rightarrow \mathcal{P}(O)$ maps
each event to a set of related objects; $\pi_{vmap} : E \rightarrow (\univ{att} \not\rightarrow \univ{val})$ maps each event to an attribute map (associating a name to a value);
$\pi_{ovmap} : O \rightarrow (\univ{att} \not\rightarrow \univ{val})$ maps each object to an attribute map;
${<}$ defines a total order on the events.
\end{definition}

The total order ${<}$ is based on the timestamp and the lexicographic order between the event identifiers.

\begin{definition}[Auxiliary Object-Centric Definitions]\label{def:auxiliaryDefinitions}
Given an object-centric event log $L = (E, \allowbreak O, \allowbreak \pi_{otyp}, \allowbreak \pi_{act}, \allowbreak \pi_{time}, \allowbreak \pi_{omap}, \allowbreak \pi_{vmap}, \allowbreak \pi_{ovmap}, \allowbreak {<})$, we define:
\begin{itemize}
\item {\bf Lifecycle of an Object} $\textrm{lif} : O \rightarrow \mathcal{P}(E)$, $\textrm{lif}(o) = \{ e \in E ~ \arrowvert ~ o \in \pi_{omap}(e) \}$
\item {\bf Start/End events for the Lifecycle of an Object} $\textrm{start}(o) = \textrm{argmin}_{{<}} \textrm{lif(o)}$, $\textrm{end}(o) = \textrm{argmax}_{{<}} \textrm{lif(o)}$
\item {\bf Eventually-Follows Graph for an Object} $\textrm{efg} : O \rightarrow \mathcal{P}(E \times E)$, $\textrm{efg}(o) = \{ (e_1, e_2) \in \textrm{lif}(o) \times \textrm{lif}(o) ~ \arrowvert ~ e_1 {<} e_2 \}$.
\item {\bf Directly-Follows Graph for an Object} $\textrm{dfg} : O \rightarrow \mathcal{P}(E \times E)$, $\textrm{dfg}(o) = \{ (e_1, e_2) \in \textrm{efg}(o) ~ \arrowvert ~ \not\exists_{e_3, o \in \pi_{omap}(e_3)} ~ e_1 {<} e_3 {<} e_2 \}$.
\item {\bf Objects of a given Object Type} For any $ot \in \univ{OT}$, $O_{ot} = \{ o \in O ~ \arrowvert ~ \pi_{otyp}(o) = ot \}$
\item {\bf Objects Interaction} $\textrm{interact} : O \rightarrow \mathcal{P}(O)$, $\textrm{interact}(o) = \{ o' \in O ~ \arrowvert ~ \exists_{e \in E} ~ o \in \pi_{omap}(e) ~ \wedge ~ o' \in \pi_{omap}(e) \}$.
For any $ot \in \univ{OT}$, $\textrm{interact}_{ot} : O \rightarrow \mathcal{P}(O_{ot})$, $\textrm{interact}_{ot}(o) = \{ o' \in \textrm{interact}(o) ~ \arrowvert ~ \pi_{otyp}(o') = ot \}$.
\item {\bf Objects Creation} For any $ot \in \univ{OT}$, $\textrm{creation}_{ot} : O \rightarrow \mathcal{P}(O_{ot})$, $\textrm{creation}_{ot}(o) = \{ o' \in \textrm{interact}_{ot}(o) ~ \arrowvert ~ \pi_{time}(\textrm{start}(o)) < \pi_{time}(\textrm{start}(o')) \}$.
\item {\bf Objects Continuation} For any $ot \in \univ{OT}$, $\textrm{continuation}_{ot} : O \rightarrow \mathcal{P}(O_{ot})$, \\ $\textrm{continuation}_{ot}(o) = \{ o' \in \textrm{interact}_{ot}(o) ~ \arrowvert ~ \pi_{time}(\textrm{end}(o)) = \pi_{time}(\textrm{start}(o')) \}$.
\item {\bf Objects Co-birth} For any $ot \in \univ{OT}$, $\textrm{cobirth}_{ot} : O \rightarrow \mathcal{P}(O_{ot})$, $\textrm{cobirth}_{ot}(o) = \{ o' \in \textrm{interact}_{ot}(o) ~ \arrowvert ~ \pi_{time}(\textrm{start}(o)) = \pi_{time}(\textrm{start}(o')) \}$.
\item {\bf Objects Co-death} For any $ot \in \univ{OT}$, $\textrm{codeath}_{ot} : O \rightarrow \mathcal{P}(O_{ot})$, $\textrm{codeath}_{ot}(o) = \{ o' \in \textrm{interact}_{ot}(o) ~ \arrowvert ~ \pi_{time}(\textrm{end}(o)) = \pi_{time}(\textrm{end}(o')) \}$.
\item {\bf Common Attributes for the Objects of a given Object Type} For any $ot \in \univ{OT}$, $\textrm{OATT}_{ot} = \{ a \in \univ{att} ~ \arrowvert ~ a \in \textrm{dom}(\pi_{ovmap}(o)) ~ \forall o \in O_{ot} \}$
%\item For any $ot \in \univ{OT}$, $\textrm{OVAL}_{ot} = \{ \pi_{ovmap}(o)(a) ~ \arrowvert ~ o \in O_{ot} ~ \wedge ~ a \in \textrm{OATT}_{ot} \}$
\end{itemize}
\end{definition}

Definition \ref{def:auxiliaryDefinitions} outlines key concepts such as \emph{lifecycle} and establishes the directly- and eventually-follows graph. It also forms associations through various object interactions, initially discussed in \cite{berti2022graph}. An example is the object co-birth graph, linking objects that start their lifecycle simultaneously. In the case study \cite{berti2023analyzing}, these interactions help identify objects with incomplete lifecycles and detect orders lacking associated payments. Additionally, we define $\textrm{OATT}_{ot}$ as the set of attributes applicable to all objects of a specific type.

\subsection{Object-Centric Feature Maps}
\label{subsec:objCentrFeaMaps}

To apply machine learning algorithms to object-centric event logs, we need to convert them to a set of numerical features. To extract such numerical features, we report the methodology introduced in \cite{berti2022graph}.

\begin{definition}[Object-Centric Feature Map]\label{def:objCentrFeaMap}
Given an object-centric event log $L = (E, \allowbreak O, \allowbreak \pi_{otyp}, \allowbreak \pi_{act}, \allowbreak \pi_{time}, \allowbreak \pi_{omap}, \allowbreak \pi_{vmap}, \allowbreak \pi_{ovmap}, \allowbreak {<})$, an object type $ot \in \univ{OT}$,
and a set of strings $\Sigma \subseteq \univ{\Sigma}$, a feature map is a function $O_{ot} \rightarrow (\Sigma \rightarrow \mathbb{R})$.
\end{definition}

First, we introduce in Definition \ref{def:objCentrFeaMap} a generic definition of object-centric feature map. Then, in Definition \ref{def:exampleObjCentrFeaMap}, we introduce an example object-centric feature map computed using the definitions introduced in Definition \ref{def:auxiliaryDefinitions}.

\begin{definition}[Example of Object-Centric Feature Map]\label{def:exampleObjCentrFeaMap}
Let $+$ be the string concatenation operator.
Given an object-centric event log $L = (E, \allowbreak O, \allowbreak \pi_{otyp}, \allowbreak \pi_{act}, \allowbreak \pi_{time}, \allowbreak \pi_{omap}, \allowbreak \pi_{vmap}, \allowbreak \pi_{ovmap}, \allowbreak {<})$ and an object type $ot \in \univ{OT}$,
we define $F_{ot} : O_{ot} \rightarrow (\Sigma \rightarrow \mathbb{R})$ such that:
\begin{itemize}
\item {\bf Numeric Attribute Values} For any $\textrm{att} \in \textrm{OATT}_{ot}$, if $v = \pi_{ovmap}(o)(\textrm{att}) \in \mathbb{R}$, then $F_{ot}(o)(\textrm{``numvalue''}+\textrm{att}) = v$
\item {\bf One-Hot Encoding of String Attribute Values} For any $\textrm{att} \in \textrm{OATT}_{ot}$, if $v = \pi_{ovmap}(\textrm{att}) \in \univ{\Sigma}$, then  $F_{ot}(o)(\textrm{``strvalue''}+\textrm{att}+\textrm{``\_''}+v) = 1$
\item {\bf Count of the Activities} For any $a \in \univ{act}$, $F_{ot}(o)(\textrm{``lifecyclecontains''}+a) = \arrowvert \{ e \in \textrm{lif}(o) ~ \arrowvert ~ \pi_{act}(e) = a \} \arrowvert$
\item {\bf One-Hot Encoding of the Start Activities} For any $a \in \univ{act}$, $F_{ot}(o)(\textrm{``lifecyclestartswith''}+a) = \mathbbm{1}_{a = \pi_{act}(\textrm{start}(o))}$
\item {\bf Lifecycle Start Time} $F_{ot}(o)(\textrm{``lifecyclestarttime''}) = \pi_{time}(\textrm{start}(o))$
\item {\bf Lifecycle End Time} $F_{ot}(o)(\textrm{``lifecycleendtime''}) = \pi_{time}(\textrm{end}(o))$
\item {\bf Lifecycle Duration} $F_{ot}(o)(\textrm{``lifecycleduration''}) = \pi_{time}(\textrm{end}(o)) - \pi_{time}(\textrm{start}(o))$
\item {\bf Directly Follows Graph} For any $a_1, a_2 \in \univ{act}$, $F_{ot}(o)(\textrm{``dfg\_''}+a_1+\textrm{``\_''}+a_2) = \arrowvert \{ (e_1, e_2) \in \textrm{dfg}(o) ~ \arrowvert ~ \pi_{act}(e_1) = a_1 ~ \wedge ~ \pi_{act}(e_2) = a_2 \} \arrowvert$
\item {\bf Number of Interactions for a given Object Type} For any $ot' \in \univ{OT}$, $F_{ot}(o)(\textrm{``interactions''}+ot') = \arrowvert \textrm{interact}_{ot'}(o) \arrowvert$
\item {\bf Number of Creations for a given Object Type} For any $ot' \in \univ{OT}$, $F_{ot}(o)(\textrm{``creation''}+ot') = \arrowvert \textrm{creation}_{ot'}(o) \arrowvert$
\end{itemize}
We assume $\Sigma$ to contain at least the aforementioned features.
\end{definition}

In Definition \ref{def:exampleObjCentrFeaMap}, we distinguish between features related to the \emph{object attributes}, features related to the \emph{lifecycle of an object}, and \emph{features related to the interactions between the objects}.
However, the features of neighboring objects are still not exploited. An approach to resolve such limitation is proposed in Definition \ref{def:featurePropagation}.

\begin{definition}[Feature Propagation]\label{def:featurePropagation}
Let $+$ be the string concatenation operator.
Given an object-centric event log $L = (E, \allowbreak O, \allowbreak \pi_{otyp}, \allowbreak \pi_{act}, \allowbreak \pi_{time}, \allowbreak \pi_{omap}, \allowbreak \pi_{vmap}, \allowbreak \pi_{ovmap}, \allowbreak {<})$ two object types $ot, ot' \in \univ{OT}$,
and two feature maps $F_{ot} : O_{ot} \rightarrow (\Sigma_1 \rightarrow \mathbb{R})$, $F_{ot'} : O_{ot'} \rightarrow (\Sigma_2 \rightarrow \mathbb{R})$,
we define a propagated feature map $F'_{ot, agg} : O_{ot} \rightarrow (\Sigma_3 \rightarrow \mathbb{R})$ where:
\begin{itemize}
\item $\textrm{agg} : \mathcal{B}(\mathbb{R}) \rightarrow \mathbb{R}$ is an aggregation function (for instance, the mean or the median).
\item $\Sigma_3 = \Sigma_1 \cup \{ \textrm{``prop''} + \sigma_2 ~ \arrowvert ~ \sigma \in \Sigma_2 \}$
\item For any $o \in O_{ot}$ and $\sigma_1 \in \Sigma_1$, $F'_{ot, agg}(o)(\sigma_1) = F_{ot}(o)(\sigma_1)$
\item For any $o \in O_{ot}$ and $\sigma_2 \in \Sigma_2$, $F'_{ot, agg}(o)(\textrm{``prop''}+\sigma_2) = \textrm{agg}(\{ F_{ot'}(o')(\sigma_2) ~ \arrowvert ~ o' \in \textrm{interact}_{ot'}(o) \})$
\end{itemize}
\end{definition}

Definition \ref{def:featurePropagation} allows merging two feature maps, considering both the features of an object (attributes+lifecycle+interactions) and an aggregation of the features
of the neighboring objects.

\section{Approach}
\label{sec:approach}

The section detail three methods to identify anomalous feature values. Section \ref{susec:anomFeaturesViaOracles} uses the feature map and domain knowledge for anomaly recognition (methodology {\bf AF1}). Section \ref{subsec:objCentrAnomalyDetection} applies anomaly detection algorithms to score objects and examines those with the lowest scores (methodology {\bf AF2}). Section \ref{subsec:aggregatingAnomalousScores} aggregates these scores at the object level to assign scores to features (methodology {\bf AF3}).

\begin{figure*}[!b]
\vspace{-5mm}
\begin{minipage}{.49\textwidth}
\begin{lstlisting}[caption=Anomalies spotted by an LLM starting from the feature table ({\bf AF1}), frame=single, label=lst:anomaliesAF1]
1. **High Duration of Lifecycle**: The range of lifecycle durations is broad, with the maximum value at 722,032,120 while the 75% quantile is 1,215,387, indicating potential abnormally long lifecycles.
2. **Activity Occurrence Consistencies**: "Submit Purchase Order for Approval" and "Approve Purchase Order" activities typically occur once per object but have outliers with up to 100 occurrences. Investigate objects where these activities are repeated excessively.
3. **Degree of Centrality**: There's a significant jump in degree centrality values from the 75% quantile to the maximum. High centrality objects, potential interaction hubs, warrant further investigation.
4. **Rare Activities**: Activities like "Reject Purchase Order" and "(Re)Open Purchase Order" are infrequent. Focusing on these can reveal anomalies.
5. **Object Interactions**: Objects with an unusually high number of interactions in the graph may indicate anomalies.
6. **Lifecycle Start and Completion Time**: Analyze objects with exceptionally short or long lifecycles compared to the dataset trend.
7. **Objects Starting/Ending Lifecycle Together**: Examine cases where a notably high number of objects start or end their lifecycle simultaneously with the current object.
\end{lstlisting}
\end{minipage}
\hfill % Adds horizontal space between the minipages
\begin{minipage}{.47\textwidth}
\begin{lstlisting}[caption=Anomalies spotted by an LLM on the lifecycle of a single object ({\bf AF2})., frame=single, label=lst:anomaliesAF2]
1. **Duplicate Time Stamps**: Both the "Approve Requisition" and "Create Purchase Order" events are recorded at the exact timestamp of 2010-10-06 03:44:22, suggesting an immediate creation of the PO after requisition approval, which is atypical due to the expected minor delay between these actions. Similarly, "Submit Purchase Order for Approval" and "Approve Purchase Order" share the timestamp of 2010-10-06 03:45:41, indicating unusual instant approval.
2. **Order of Events**: "Approve Purchase Order" is followed by another "Submit Purchase Order for Approval" event at the same timestamp, suggesting redundancy and possibly a recording error or system glitch.
3. **Long Lifecycle Duration**: The lifecycle of PO_277871 extends unusually from 2010-10-06 to 2023-07-12, closed initially on 2010-10-08 and then reopened 13 years later, which deviates from standard P2P process durations.
4. **Close and Reopen of PO**: PO_277871 was closed on 2010-10-08 and reopened on 2023-07-12, a rare occurrence that may require verification with system administrators to understand if it reflects actual procedural needs or system setup anomalies.
\end{lstlisting}
\end{minipage}
\vspace{-5mm}
\end{figure*}

\subsection{Anomalous Features Identification through Oracles (AF1)}
\label{susec:anomFeaturesViaOracles}

Starting from an object-centric feature map, we can already get useful insights about anomalous values. A domain knowledge owner, simply looking at the feature map and its values, could spot a list of anomalous patterns in the process according to the domain knowledge (methodology {\bf AF1}). Listing \ref{lst:anomaliesAF1} shows an example detection of patterns.

\begin{figure*}[!b]
\vspace{-6mm}
\begin{minipage}{\linewidth}
    \begin{minipage}{0.37\linewidth}
        \centering
        \resizebox{\linewidth}{!}{
\begin{tabular}{lrr}
\hline
Object ID & Isolation Forest Scores & Local Outlier Factor Scores \\
\hline
PO\_23667 & -0.200785 & -40.049412 \\
PO\_23507 & -0.200311 & -7.200163 \\
PO\_23508 & -0.200311 & -7.200163 \\
PO\_23512 & -0.200311 & -7.200163 \\
PO\_23513 & -0.200311 & -7.200163 \\
PO\_23514 & -0.200311 & -7.200163 \\
PO\_23515 & -0.200311 & -7.200163 \\
PO\_23516 & -0.200311 & -7.200163 \\
PO\_23517 & -0.200311 & -7.200163 \\
PO\_277871 & -0.195874 & -7.622763 \\
PO\_23511 & -0.189318 & -7.200163 \\
PO\_3903 & -0.187092 & -54.929239 \\
PO\_133097 & -0.175762 & -8.086049 \\
PO\_23668 & -0.174838 & -39.503084 \\
PO\_23669 & -0.174838 & -39.503084 \\
PO\_23510 & -0.174382 & -7.217331 \\
PO\_86355 & -0.172010 & -3.117746 \\
PO\_85465 & -0.171363 & -0.125512 \\
PO\_23518 & -0.170136 & -7.212333 \\
PO\_23519 & -0.170136 & -7.212333 \\
PO\_23520 & -0.170136 & -7.212333 \\
PO\_23521 & -0.170136 & -7.212333 \\
PO\_23522 & -0.170136 & -7.212333 \\
PO\_84184 & -0.169095 & -1.549233 \\
PO\_3836 & -0.168964 & -213.993317 \\
PO\_3837 & -0.168964 & -213.982787 \\
PO\_3838 & -0.168964 & -213.974041 \\
PO\_3839 & -0.168964 & -213.967588 \\
PO\_3840 & -0.168964 & -213.960370 \\
PO\_3841 & -0.168964 & -213.953323 \\
\hline
\end{tabular}
        }
        \captionof{table}{Anomaly scores for some purchase orders of the considered log.}
		\label{tab:anomalyScoreSomePurchaseOrders}
    \end{minipage}
    \hfill
    \begin{minipage}{0.5\linewidth}
        \centering
        \resizebox{\linewidth}{!}{
\begin{tabular}{|p{7cm}|c|c|}
\hline
\textbf{Feature (with Value)} & \textbf{Count} & $\textrm{FEA\_SCORE}$ \\ 
\hline
1 Occurrence of the activity Cancel Purchase Order & 300 & -0.07  \\
\hline
1 Occurrence of the activity (Re)Open Purchase Order & 167 & -0.12  \\
\hline
44 other orders are terminating with the same event & 45 & -0.21  \\
\hline
45 other objects are interacting with the order & 45 & -0.21  \\
\hline
The activity Approve Purchase Order is not executed & 131 & -0.07  \\
\hline
There are 2 activities in the lifecycle of the order & 72 & -0.09  \\
\hline
29 other orders are terminating with the same event & 30 & -0.19  \\
\hline
30 other objects are interacting with the order & 30 & -0.19  \\
\hline
27 other orders are terminating with the same event & 28 & -0.19  \\
\hline
28 other objects are interacting with the order & 28 & -0.19  \\
\hline
20 other orders are terminating with the same event & 21 & -0.18  \\
\hline
There is a single event in the lifecycle of the order & 53 & -0.04  \\
\hline
The activity Submit Purchase Order for Approval is not executed & 53 & -0.04  \\
\hline
There are 13 events in the lifecycle of the order & 41 & -0.05  \\
\hline
\end{tabular}
        }
        \captionof{table}{Features' values correlated with anomalies (methodology {\bf AF3}).}
		\label{tab:featuresValuesCorrelatedAnomalies}
    \end{minipage}
\end{minipage}
\vspace{-3mm}
\end{figure*}

\begin{definition}[Features' Oracle]\label{def:featuresOracle}
Given an object-centric event log $L = (E, \allowbreak O, \allowbreak \pi_{otyp}, \allowbreak \pi_{act}, \allowbreak \pi_{time}, \allowbreak \pi_{omap}, \allowbreak \pi_{vmap}, \allowbreak \pi_{ovmap}, \allowbreak {<})$, an object type $ot \in \univ{OT}$,
and a feature map $F_{ot} : O_{ot} \rightarrow (\Sigma \rightarrow \mathbb{R})$, we define as oracle any function $\textrm{ORACLE}_{\Sigma} : \Sigma \rightarrow (\mathbb{R} \rightarrow \mathbb{R})$
\end{definition}

Definition \ref{def:featuresOracle} formally introduces an ``oracle'' function looking at the values of the feature map and associating them with a real number. In our setting, we can assume that the oracle associates negative real numbers with anomalous feature values.

\subsection{Scoring the Anomalousness of an Object (AF2)}
\label{subsec:objCentrAnomalyDetection}

Having an object-centric feature map, we could apply any anomaly detection algorithm (such as isolation forests or local outlier factor) to assign an anomaly score to the objects. The objects having lower anomaly score are considered anomalous.

\begin{definition}[Objects' Score Function]\label{def:objectsScoreFunction}
Given an object-centric event log $L = (E, \allowbreak O, \allowbreak \pi_{otyp}, \allowbreak \pi_{act}, \allowbreak \pi_{time}, \allowbreak \pi_{omap}, \allowbreak \pi_{vmap}, \allowbreak \pi_{ovmap}, \allowbreak {<})$ and an object type $ot \in \univ{OT}$,
we define as score function any function $\textrm{SCORE}_{ot} : O_{ot} \rightarrow \mathbb{R}$.
\end{definition}

In Definition \ref{def:objectsScoreFunction}, we formally introduce a score function associating each object with a real number. In our setting, the score function is the anomaly detection algorithm.

\begin{definition}[Objects' Score Rank Function]\label{def:objectScoreRankFunction}
Given an object-centric event log $L = (E, \allowbreak O, \allowbreak \pi_{otyp}, \allowbreak \pi_{act}, \allowbreak \pi_{time}, \allowbreak \pi_{omap}, \allowbreak \pi_{vmap}, \allowbreak \pi_{ovmap}, \allowbreak {<})$, an object type $ot \in \univ{OT}$,
and a score function $\textrm{SCORE}_{ot} : O_{ot} \rightarrow \mathbb{R}$, we define as rank any injective function
$\textrm{RANK}_{\textrm{SCORE}_{ot}} : O_{ot} \rightarrow \mathbb{N}$ such that for any $o_1, o_2 \in O, o_1 \neq o_2$:
$$\textrm{RANK}_{\textrm{SCORE}_{ot}}(o_1) {<} \textrm{RANK}_{\textrm{SCORE}_{ot}}(o_2) ~ \iff ~ \textrm{SCORE}_{ot}(o_1) {<} \textrm{SCORE}_{ot}(o_2)$$
\end{definition}

The score (for instance, related to the application of an anomaly detection algorithm) is used in Definition \ref{def:objectScoreRankFunction} to introduce a rank between the objects.
In methodology {\bf AF2}, we propose to explore the lifecycle of the most anomalous objects with the goal of understanding the anomalous patterns. Listing \ref{lst:anomaliesAF2} shows an example detection of patterns on the lifecycle of an object.

\subsection{Anomalous Features Identification aggregating Anomaly Scores}
\label{subsec:aggregatingAnomalousScores}

Our methodology {\bf AF3} aims to use the anomaly scores at the object level to automatically assign an anomaly score to the values of the feature map.

\begin{definition}[Normalization of a Feature Map]\label{def:normalizationFeatureMap}
Given an object-centric event log $L = (E, \allowbreak O, \allowbreak \pi_{otyp}, \allowbreak \pi_{act}, \allowbreak \pi_{time}, \allowbreak \pi_{omap}, \allowbreak \pi_{vmap}, \allowbreak \pi_{ovmap}, \allowbreak {<})$, an object type $ot \in \univ{OT}$,
a feature map $F_{ot} : O_{ot} \rightarrow (\Sigma \rightarrow \mathbb{R})$, and a real number $\epsilon > 0$, we define for $\sigma \in \Sigma$:
\begin{itemize}
\item $\textrm{min}_{\sigma} = \textrm{min} \{ F_{ot}(o)(\sigma) ~ \arrowvert ~ o \in O_{ot} \}$.
\item $\textrm{max}_{\sigma} = \textrm{max} \{ F_{ot}(o)(\sigma) ~ \arrowvert ~ o \in O_{ot} \}$.
\item $F^{norm}_{ot, \epsilon} : O_{ot} \rightarrow ( \Sigma \rightarrow [-1, 1] )$,
$$F^{norm}_{ot, \epsilon}(o)(\sigma) = -1 + 2 \times \frac{F_{ot}(o)(\sigma) - \textrm{min}_{\sigma}}{\textrm{max}_{\sigma} - \textrm{min}_{\sigma} + \epsilon}$$ 
\end{itemize}
\end{definition}

Since features' values are heterogenous (for instance, the lifecycle of the objects is measured in seconds, while one-hot encoding features are valued  $0$ or $1$), we propose in Definition \ref{def:normalizationFeatureMap} an approach to normalize such values
between $-1$ and $1$.

\begin{definition}[Features' Score]\label{def:featuresScore}
Given an object-centric event log $L = (E, \allowbreak O, \allowbreak \pi_{otyp}, \allowbreak \pi_{act}, \allowbreak \pi_{time}, \allowbreak \pi_{omap}, \allowbreak \pi_{vmap}, \allowbreak \pi_{ovmap}, \allowbreak {<})$, an object type $ot \in \univ{OT}$,
a normalized feature map $F^{norm}_{ot, \epsilon} : O_{ot} \rightarrow (\Sigma \rightarrow \mathbb{R})$ and a score function $\textrm{SCORE}_{ot} : O_{ot} \rightarrow \mathbb{R}$,
we define $\textrm{FEA\_SCORE} : \Sigma \rightarrow \mathbb{R}$ such that for $\sigma \in \Sigma$
$$\textrm{FEA\_SCORE}(\sigma) = \sum_{o \in O_{ot}} \frac{\textrm{SCORE}_{ot}(o) \times F^{norm}_{ot, \epsilon}(o)(\sigma)}{\arrowvert O_{ot} \arrowvert}$$
\end{definition}

Definition \ref{def:featuresScore} implements {\bf AF3} by assigning a score to each feature starting from the scores at the object level. Table \ref{tab:featuresValuesCorrelatedAnomalies} shows an example in which the objects' anomaly scores are aggregated to provide the most anomalous features' values.

\section{Case Study}
\label{sec:discussion}

In this section, we discuss the application of the proposed techniques on top of a real-life P2P object-centric event log (ECE group).

\subsection{Context}

ECE began using the Celonis platform for process mining in 2020, integrating systems like xFlow for document acquisition and SAP ERP. Traditional process mining faced issues such as convergence and divergence \cite{DBLP:conf/sefm/Aalst19}. Initially, ECE used the multi-event log approach in Celonis, but later transitioned to tools and libraries from the PADS group at RWTH Aachen University. Insights from this case study were detailed in \cite{berti2023analyzing} and shared with stakeholders through seminars.

We are interested in applying anomaly detection to discover deviations from the expected behavior (non-compliance, such as \emph{maverick buying}, i.e. inserting formally the order only after its placement, and \emph{post-mortem changes to purchase requisitions}) and identify behavior leading to a monetary loss in the P2P process (for example,
invoice paid double, or discount rates not taken because of invoices taking long to process, or non-justified payment blocks). The aforementioned non-compliant and/or non-optimal behavior could in principle be identified with conformance checking techniques rather than anomaly detection.
However, that requires pinpointing a priori all the causes of deviations. Anomaly detection can instead be applied without requiring prior knowledge of the possible deviations.

Our analysis primarily utilized the \emph{pm4py} process mining library \cite{DBLP:journals/simpa/BertiZS23} and the \emph{OC-PM} Javascript-based tool \cite{DBLP:journals/sttt/BertiA23}, which both support object-centric feature extraction as outlined in \cite{berti2022graph}. In previous work, we used these tools in a case study \cite{berti2023analyzing}. pm4py provides a dataframe via $\mathtt{pm4py.extract\_ocel\_features}$, compatible with any Python machine learning library. OC-PM\footnote{\url{https://www.ocpm.info/}}, after feature extraction, employs the ``Isolation Forests'' anomaly detection algorithm.

\subsection{Methodologies and Algorithms}

By applying the methodologies {\bf AF1}, {\bf AF2}, and {\bf AF3}, we can identify some inherent differences.
The method in {\bf AF1} bypasses anomaly detection algorithms, reducing computational costs and domain knowledge requirements, but only evaluates single features, not their combinations. {\bf AF2} can identify anomalies across feature combinations but requires extensive domain knowledge exploration of object lifecycles, demanding more time. {\bf AF3} operates without domain knowledge, potentially resulting in a lengthy list of anomalous features that may challenge analysts.

The methods used in the analysis of object-centric features, including feature selection, dimensionality reduction, and anomaly detection, were selected among the most popular options. For feature selection, variance was used to retain features with significant variability, suggesting their importance in distinguishing between data points. Dimensionality reduction employed \emph{Principal Component Analysis (PCA)} and \emph{FastMap}, both effective in reducing the number of variables. PCA transforms data into principal components, linear combinations of the original variables, while FastMap is a distance-preserving projection that maps data into a lower-dimensional space. Anomaly detection involved \emph{Isolation Forests} and \emph{Local Outlier Factor (LOF)}, chosen for their ability to identify outliers. Isolation Forests isolate anomalies by selecting a feature and a split value randomly, while LOF measures local deviation of data points from their neighbors to identify similar density regions.

In our P2P object-centric setting, FastMap was the preferred method due to its ability to maintain non-linear relationships and computational efficiency. Unlike PCA, which involves intensive eigen-decomposition and can be less suitable for large datasets with ambiguous component interpretations, FastMap efficiently reduces high-dimensional data into lower dimensions without requiring full distance matrix computations. This is particularly advantageous for managing graph-based features.

The findings highlight the strengths of Isolation Forests and LOF in anomaly detection. Isolation Forests are effective for high-dimensional data and large volumes, isolating anomalies using decision tree splittings without needing pairwise distance calculations. This accelerates anomaly detection in complex datasets. LOF excels at identifying anomalies in specific subgroups by calculating local density deviations, useful for clustered data. However, LOF requires more computational resources for large datasets. In our analysis, Isolation Forests successfully detect anomalies in object-centric event logs with traditional lifecycle features, while LOF is preferable for graph-based features, focusing on local context to identify anomalies in networks of object interactions.

\subsection{Refinement of the Analysis}

\begin{figure*}[!t]
\centering
\includegraphics[width=0.75\textwidth]{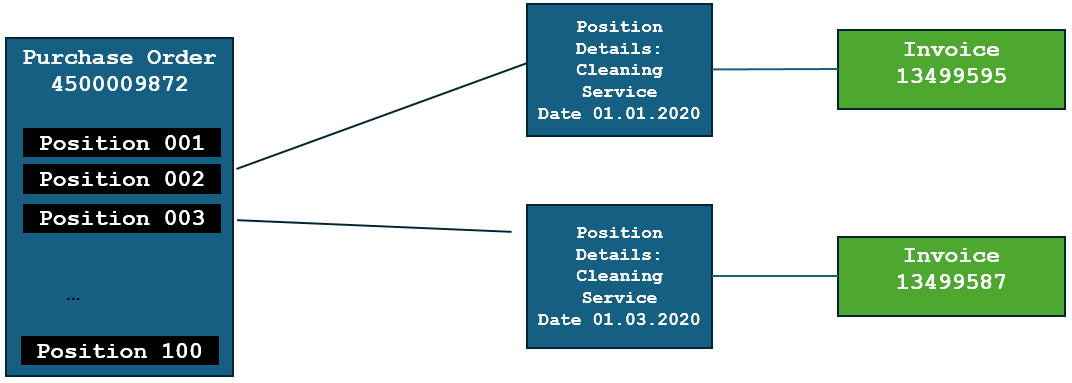}
\caption{Interaction between maintenance contracts with several positions and invoices.}
\label{fig:necessFeaPropaga}
\vspace{-4mm}
\end{figure*}

After performing an initial analysis, we performed some postprocessing of the object-centric event log and applied feature propagation to enhance the results.

We have hundreds of activities in the object-centric event log, mostly related to changing field values (change tables in SAP). Most of them are not relevant for object-centric anomaly detection and increase the dimensionality of the data with little gain.
After our first application of anomaly detection, we repeated it on an object-centric event log that was filtered keeping only the relevant activities. The selection of relevant activities proved challenging on its own.
Some infrequent activities, which were the first candidates for removal, identify important anomalies. We could distinguish between manual and automatic activities, with the latter being less important for anomaly detection.

We discovered that a traditional object-centric feature map based on the lifecycle and interactions of object types gives an incomplete process view. For instance, we found that invoices were often blocked for orders lacking preliminary purchase requisition approval, a pattern not visible when considering only invoices. By extending invoice data with information from related purchase orders (using Definition \ref{def:featurePropagation}), we identified the root cause of this performance issue. Another observation, illustrated in Figure \ref{fig:necessFeaPropaga}, showed that orders with multiple positions (e.g., maintenance contracts) might appear anomalous when viewed in isolation. However, considering each item's direct relation to an invoice, such behavior is not anomalous.

\subsection{Main Results}

Anomaly detection allowed us to identify several non-compliance issues in the P2P process.
We identified a non-negligible amount of orders with the \emph{maverick buying} problem.
The order is placed to the supplier skipping all the approval steps, the supplier sends an invoice to the company,
and only then the purchase order is formally created in the ERP system.
Moreover, we recorded several change activities done to purchase requisitions after their approval in order to match the amounts/quantities of the purchase order (\emph{post-mortem changes to PRs}). This is a deleterious behavior as the purchase requisition was deliberately proposed to the managers with a lower amount.

Looking at the inefficiencies in the process leading to a monetary loss, we observed orders invoiced (and paid) several times, which were not maintenance contracts.
Moreover, we identified invoices with an excessive number of change activities, signaling an inefficiency in the process (as this behavior is correlated with longer processing times).
Considering the interaction between purchase orders, invoices, and payments, we observed that inefficiencies in the purchase orders also lead to inefficient processing of payments.

\subsection{Limitations of LLMs as Domain Knowledge Providers}

We used LLMs to interpret results, following methods in \cite{DBLP:conf/bpm/Berti0A23}. Specifically, \\ $\mathtt{pm4py.llm.abstract\_ocel\_features}$ was used for textual abstraction in method {\bf AF1}, and $\mathtt{pm4py.llm.abstract\_ocel}$ for {\bf AF3}. The \emph{gpt-4-turbo} LLM model, available as of \emph{09-04-2024} in Germany, was chosen for its large context window to generate insights.

Applying LLMs to textual abstractions from our object-centric event log produced mixed results. For methodology {\bf AF1}, the insights helped identify anomalous patterns and filter objects for further analysis using the OC-PM tool.

However, several limitations arose. The context window of the LLM, despite improvements with the \emph{gpt-4-turbo} model, restricted the inclusion of lifecycles with many events, limiting the application of methodology {\bf AF2} to objects with fewer events. Inconsistencies across different sessions were noted \cite{DBLP:conf/bpm/Berti0A24}, sometimes requiring the merging of insights from different sessions as an "ensemble". Hallucinations and irrelevant outputs compared to the original prompt also occurred \cite{DBLP:conf/bpm/Berti0A24}.

\section{Conclusion}
\label{sec:conclusion}

In this paper, we explored methodologies for object-centric anomaly detection and their implementation challenges in a real-life P2P process. We discovered that selecting appropriate algorithms is crucial, as they vary in effectiveness depending on the type of object-centric features. Identifying anomalous feature values is only part of the process; interpreting these results typically requires domain knowledge. Large Language Models (LLMs) can supplement domain knowledge, enabling analysts with limited process expertise to gain insights. However, challenges such as hallucinations and output inconsistency in LLMs must be noted.

\bibliographystyle{splncs04}
\bibliography{references}

\end{document}